\documentclass{aa}  
\usepackage{graphicx}
\usepackage{amsmath}
\usepackage{hyperref}
\usepackage[capitalize]{cleveref}
\usepackage{txfonts}
\usepackage{natbib}
\bibpunct{(}{)}{;}{a}{}{,} 
\usepackage{color}
\usepackage[dvipsnames]{xcolor}
\usepackage{xspace}

\renewcommand{\rho}{\varrho}
\newcommand{\PD}[2]{\frac{\partial#1}{\partial#2}}
\newcommand{\D}[2]{\frac{\text{d}#1}{\text{d}#2}}

\newcommand{\nabad}{\ensuremath{\nabla_\text{ad}}}
\newcommand{\HP}{\ensuremath{H_P}}
\newcommand{\Ri}{\ensuremath{\mathit{Ri}}}
\newcommand{\Ric}{\ensuremath{\mathit{Ri}_\text{c}}}
\newcommand{\Pe}{\ensuremath{\mathit{Pe}}}
\newcommand{\BVF}{Brunt--V\"ais\"al\"a frequency\xspace}

\begin{document}

\title{Testing a one-dimensional prescription of dynamical shear mixing with
a two-dimensional hydrodynamic simulation}
\titlerunning{Testing a 1D prescription of dynamical shear mixing with
a 2D hydrodynamic simulation}


\author{P.~V.~F.~Edelmann\inst{\ref{inst1}} \and F.~K.~R\"opke\inst{\ref{inst1},\ref{inst2}} \and R.~Hirschi\inst{\ref{inst3},\ref{inst4},\ref{inst5}} \and
C.~Georgy\inst{\ref{inst6},\ref{inst3}} \and S. Jones \inst{\ref{inst1}}}

\institute{Heidelberger Institut f\"ur Theoretische Studien, Schloss-Wolfsbrunnenweg 35, D-69118 Heidelberg, Germany\\
  \email{philipp@slh-code.org}\label{inst1}
  \and Zentrum f\"ur Astronomie der Universit\"at Heidelberg, Institut f\"ur Theoretische Astrophysik, Philosophenweg 12, D-69120 Heidelberg, Germany\label{inst2}
  \and Astrophysics group, Keele University, Lennard-Jones Labs, Keele ST5 5BG, UK\label{inst3}
  \and Kavli Institute for the Physics and Mathematics of the Universe (WPI), University of Tokyo, 5-1-5 Kashiwanoha, 277-8583 Kashiwa, Japan\label{inst4}
  \and UK Network for Bridging Disciplines of Galactic Chemical Evolution (BRIDGCE), \url{http://www.bridgce.ac.uk/}, UK\label{inst5}
  \and Geneva Observatory, University of Geneva, Maillettes 51, CH-1290 Sauverny, Switzerland\label{inst6}
}

\date{Received xxxx xx, xxxx / accepted xxxx xx, xxxx}

 
\abstract
{The treatment of mixing processes is still one of the major uncertainties in
1D~stellar evolution models. This is mostly due to the need to
parametrize and approximate aspects of hydrodynamics in hydrostatic codes. In
particular, the effect of hydrodynamic instabilities in rotating stars,
for example, dynamical shear instability, evades consistent description.}
{We intend to study the accuracy of the diffusion approximation to dynamical
shear in hydrostatic stellar evolution models by comparing 1D~models to a first-principle
hydrodynamics simulation starting from the same initial conditions.}
{We chose an initial model calculated with the stellar evolution code GENEC that is just
at the onset of a dynamical shear instability but does not show any other
instabilities (e.g., convection). This was mapped to the hydrodynamics code SLH
to perform a 2D~simulation in the equatorial plane.
We compare the resulting profiles in the two codes and compute an
effective diffusion coefficient for the hydro simulation.}
{Shear instabilities develop in the 2D~simulation in the regions predicted by linear theory
to become unstable in the 1D~stellar evolution model.
Angular velocity and chemical composition is redistributed in the
unstable region, thereby creating new unstable regions. After a period
of time, the system settles in a symmetric, steady state, which is
Richardson stable everywhere in the 2D~simulation, whereas the instability remains for longer in the 1D~model due to
the limitations of the current implementation in the 1D~code. A spatially resolved diffusion coefficient
is extracted by comparing the initial and final profiles of mean atomic
mass.}
{The presented simulation gives a first insight on hydrodynamics of
shear instabilities in a real stellar environment and even allows us to
directly extract an effective diffusion coefficient. We see evidence for
a critical Richardson number of 0.25 as regions above this threshold
remain stable for the course of the simulation.}


\maketitle
%

\section{Introduction}
  \label{sec:intro-motivation}
Internal mixing plays an important role in stellar evolution. Mixing
takes place under several different conditions in stellar interiors, all
linked with different kinds of fluid instability: convection, dynamical
shear, secular shear, Rayleigh-Taylor, thermohaline, etc.\ \citep[see,
for example,][]{heger2000a,maeder2013a}.
Rotation-induced instabilities (e.\,g.\ dynamical shear, secular shear,
GSF, ABCD) have a strong impact on the evolution and
structure of stars. Stellar
models including prescriptions for rotation can, depending upon
the initial angular velocity, exhibit significantly different structural
and evolutionary characteristics to non-rotating models \citep[see,
e.g.,][]{heger2000a, heger2000b, hirschi2004a, ekstroem2012a}. Most
predominantly, rotating models have more massive cores, higher
luminosities and more efficient mass loss via winds.  In massive stars,
the interior structure of the carbon-oxygen core is determined in a
large part by its mass and by mixing at convective boundaries
\citep{young2005a}. Whether or not a star is rotating (and how fast)
can, primarily by dictating the mass of the CO core and secondarily by
inducing mixing within the core, alter the pre-supernova density
profile of the star. A variation in density profile \citep[not to
mention the pre-supernova angular velocity profile, see,
e.g.,][]{fryer2000a} can have a marked effect on the explosion energy,
explosive nucleosynthetic yield and compact remnant type (neutron star
or black hole) and mass \citep[see][and references
therein]{oconnor2013a,sukhbold2014a,ertl2016a,sukhbold2016a}.

Rotation also affects the nucleosynthesis in massive stars, especially at low metallicities.
Rotation-induced mixing across convectively stable layers in massive
stars has been shown to be able to explain the observed enhancement of
nitrogen in the early universe \citep[see][and references therein]{hirschi2007a}. In the absence of
additional mixing processes, the surface nitrogen abundance should scale
with the metallicity of the protostellar gas. However, transport of
$^{12}$C and $^{16}$O from the convective He-burning across the
convectively stable He layer and into the H-burning shell by the secular
shear instability enables a primary production of $^{14}$N by proton
captures on the transported $^{12}$C and $^{16}$O. The same secular
shear process also transports some of this $^{14}$N from the H-burning
shell in which it was created back down into the convective He-burning
core, where it is exposed to higher temperatures and alpha particles,
resulting in the creation of $^{22}$Ne (two alpha-captures). $^{22}$Ne
is an efficient neutron source (activated by alpha-capture;
$^{22}\mathrm{Ne}(\alpha,n)^{25}\mathrm{Mg}$) and its
rotationally induced presence in the He-core boosts the weak slow
neutron-capture ($s$) process production at low metallicities
\citep[see][for more details]{frischknecht2012a,frischknecht2016a}.

Rotation and all the other processes mentioned above are
intrinsically multi-dimensional due to their turbulent nature, and their
implementation in one-dimensional (1D) stellar evolution codes
is not straightforward and relies on many simplifications. Usually, they
are each accounted for as separate diffusive processes, the diffusion
coefficients of which are summed up. The number of considered
instabilities and
the way they are implemented in stellar evolution codes lead to
significant differences between the results of various codes,
particularly for the advanced evolutionary stages of massive stars
\citep{martins2013a}.

The results of stellar evolution calculations are thus model-dependent, and while the impact of
rotation is qualitatively determined, there are still free
parameters and artifacts of the numerical implementations of the
rotation physics (including the artificial symmetry of rotating 1D
models) that prevent the 1D models from making precise quantitative
predictions. A better knowledge about how chemical species and angular
momentum are transported inside stars is thus a crucial step in
improving the treatment of rotation in, and hence the predictive power
of, stellar evolution codes.

Because of the considerable increase in computing power and the
development of modern and efficient hydrodynamics codes during the past
decades, it has become possible to simulate these processes by means of
multi-dimensional hydrodynamics simulations. Most of the efforts in this
direction undertaken recently have been focused on convection in the
central regions or shells \citep{herwig2006a,meakin2007a,jones2017a}, or
the convective envelope of cool stars
\citep{freytag2008a,chiavassa2009a,viallet2013b,magic2013a}. On the side
of rotation-induced instabilities, \citet{prat2013a,prat2014a} and
\citet{prat2016a} study the transport of chemicals through
three-dimensional (3D) direct numerical simulation of secular shear
mixing and compare their results to various prescriptions usually found
in 1D~stellar evolution codes. They find a qualitative agreement with
the model of \citet{zahn1992a} and extract a diffusion coefficient for
secular shear mixing from their simulations in Boussinesq and small
P\'eclet number approximation. \citet{garaud2015a} performed a stability
analysis of secular shear mixing in low-P\'eclet-number flows. In a
linear stability analysis they find that flows with $\Ri Pe \gtrsim
0.25$ are stable and that $\Ri \gtrsim Pr^{-1}$ is a criterion for
energy stability for large Reynolds numbers. In direct numerical
simulations they see instabilities developing for $\Ri Pe > 0.25$ but
disappearing significantly below the limit given by the energy analysis.
\citet{garaud2016a} continue this numerical study of shear flows for a
wider range of parameters and identify different categories of flow
depending of the value of $\Ri Pe$.

\citet{brueggen2001a} performed simulations of Kelvin--Helmholtz
instabilities in stratified shear layers. They found instability also in
the parameter range that is predicted to be stable by linear theory and
gave an expression for the effective diffusion coefficient extracted
from the simulations using tracer particles. The significance of their
result is limited by the fact that they use Boussinesq approximation to
compute the Richardson number, which is used as the criterion for
instability. This approximation is only valid for small variations in
density, a criterion not fulfilled in their simulations, where density
varies by almost 15\%.

The aim of this paper is to answer a few specific questions about
dynamical shear instabilities in stellar models by comparing the
1D~prescription to results of a 2D~hydrodynamic simulation.

\begin{enumerate}
  \item The Richardson criterion itself is only a necessary condition
    based on a simple energy argument (see \cref{sec:theory}).  Does the
    instability actually develop in a realistic stellar setting?
  \item Are the position and extent of the mixing region comparable?
  \item In regimes with comparably short time scales in the
    1D~simulation the effect of the instability is spread out over many
    time steps. The position of the unstable regions can shift because
    of the change in the profiles caused by mixing. Is a similar
    behavior observed when including hydrodynamic effects?
  \item Given that the major effect of shear mixing is changing the
    profiles of angular velocity and composition, how well do the initial
    and final state of the hydrodynamic simulation compare to the
    1D~prescription?
  \item Another commonly observed phenomenon is self-quenching of the
    instability due to the change in the angular velocity profile. Does
    this effect also appear in the full hydrodynamic description?
  \item A useful quantitative comparison between 1D and 2D/3D data is
    often hard because the 1D~prescription of mixing is typically using
    a diffusion equation, while the actual process is of advective
    nature. Does a $D$ exist that can reproduce the change in the
    radially averaged profiles of $\bar{A}$ and $\Omega$ by using only a
    diffusion approximation?
\end{enumerate}

In \cref{sec:theory}, we recall some theoretical aspects of the
dynamical shear instability.
In \cref{sec:codes}, we describe the stellar model we used as input to our
hydrodynamics simulation, and give a brief description of the Seven-League
hydrodynamics code.  \Cref{sec:model_setup} is dedicated to the mapping
of our 1D~model onto the 2D~mesh used for the hydrodynamics simulation.
In \cref{sec:hydro}, we discuss the results of our multi-dimensional
simulation of dynamical shear mixing, and in
\cref{sec:comparison-1d-2d} we compare it to the results obtained with
the 1D~stellar evolution code. Finally, our conclusions are presented in
\cref{sec:conclusions}.

\section{Theory of the dynamical shear instability}
\label{sec:theory}
In a setting without gravity, two layers of a fluid moving side-by-side
at different velocities are subject to the Kelvin--Helmholtz
instability. In the presence of gravity this is not always the case. A
simple energy argument can be invoked to derive a necessary condition
for the instability of an initially laminar flow
\citep[e.g.,][Sect.~3.6.3]{sutherland2010a}.

Consider the mixing of the fluid in two layers at distance~$\delta z$.
The difference in density between the two is $\delta \rho$ and the
difference in velocity is $\delta u$. The latter can be positive or
negative. After the mixing event both layers will move at the velocity
$u + 1/2 \delta u$ due to conservation of momentum. The difference in
density between the two states is neglected here under the Boussinesq
approximation \citep[e.g.,][Sect.~1.12]{sutherland2010a}. In essence the
approximation states that density is assumed to be spatially and
temporally constant, except when it is used for calculating buoyancy.

The new state has a lower kinetic energy than the initial one. The
difference is

\begin{equation}
  \Delta E_\text{kin} = \rho \left(u + \frac{\delta u}{2}\right)^2 -
  \left(\frac{1}{2} \rho u^2 + \frac{1}{2}\rho\left(u+\delta u\right)^2
  \right) = - \frac{1}{4} \rho \delta u^2.
\end{equation}
This excess energy needs to be enough to provide the potential energy
for displacement against the (positive) gravitational acceleration~$g$
for the exchange of the two layers given by

\begin{equation}
  \Delta E_\text{pot} = g \delta \rho \delta z.
\end{equation}
This requires $\Delta E_\text{kin} + \Delta E_\text{pot} < 0$, leading
to a necessary condition for instability
\begin{equation}
  \frac{1}{4} \geq \frac{g \delta\rho \delta
  z}{\rho (\delta u)^2}.
\end{equation}
It can also be stated in its differential form as
\begin{equation}
  \label{eq:ri-def}
  \Ri \equiv \frac{g}{\rho} \frac{\partial\rho/\partial
  z}{(\partial u / \partial z)^2} \leq \frac{1}{4}.
\end{equation}
This defines the well-known Richardson number \Ri{} in Boussinesq
approximation. The threshold 1/4 is called the critical Richardson
number~\Ric.

A rigorous proof of $\Ric=1/4$ was given by \citet{miles1961a} and later
in simpler and more general fashion by \citet{howard1961a}. They
consider the case of an inviscid, incompressible fluid with a vertical
density stratification. It is important to note that their linear
analysis starts from an initially laminar flow, that means turbulence cannot
develop through shear from an initially nonturbulent configuration.
\citet{canuto2001a} call this the \emph{bottom-up} approach. The opposite is
a \emph{top-down} approach, which asks the question whether already existing
turbulent motion is quenched by the stabilizing effect of the density gradient.
An analytical answer in Boussinesq approximation has been given by
\citet{abarbanel1984a}, who find that the necessary and sufficient condition
for stability of a shear layer is $\Ri > 1$.

It is common to rewrite the definition of \Ri{} in terms of angular
velocity~$\Omega$ and the \BVF~$N$ \citep[e.g.,][]{maeder2009a},
\begin{equation}
  \label{eq:ri-bvf}
  \Ri = \left(\frac{N}{\varpi \PD{\Omega}{r}}\right)^2.
\end{equation}
This uses the distance to the rotational axis~$\varpi= r \sin
\vartheta$, where $\vartheta$ is the colatitude.

The \BVF~$N$ of a stratified gas is given by \citep[e.g.,][Sect.\
6.4.1]{maeder2009a},
\begin{equation}
  \label{eq:bvf}
  N^2 = \frac{g \delta}{\HP} \left( \nabla_\text{int} - \nabla +
  \frac{\varphi}{\delta} \nabla_\mu\right).
\end{equation}
It is the frequency at which density perturbations in the stratification
oscillate. Terms related to rotation are not included in this definition
of $N$ as they to do not appear in the Richardson criterion
\citep[see][Sect.\ 12.2.1.1]{maeder2009a}.

Here, $g$ is the magnitude of
effective gravitational acceleration, including the centrifugal force. The
parameters $\delta$ and $\varphi$ are partial derivatives of density,
determined by the equation of state,
\begin{equation}
  \label{eq:eos-delta-phi}
  \delta = - \left(\PD{\ln \rho}{\ln T}\right)_{P,\mu}, \quad
  \varphi = \left(\PD{\ln \rho}{\ln \mu}\right)_{P,T}.
\end{equation}
The pressure scale height is defined by
\begin{equation}
  \label{eq:Hp}
  \HP = - \D{r}{P} P.
\end{equation}
There are three gradients involved in \cref{eq:bvf}. The external
gradient,
\begin{equation}
  \label{eq:nabla}
  \nabla = \D{\ln T}{\ln P},
\end{equation}
is given by the temperature stratification of the gas. As we are
considering the deep interior of a star, the internal temperature
gradient of the displaced element is equal to the adiabatic gradient,
\begin{equation}
  \label{eq:nabla-ad}
  \nabla_\text{int} = \nabla_\text{ad} = \frac{P \delta}{C_P \rho T},
\end{equation}
with the specific heat at constant pressure~$C_P$. The gradient in mean
molecular weight~$\mu$ is given by
\begin{equation}
  \label{eq:nabla-mu}
  \nabla_\mu = \D{\ln \mu}{\ln P}.
\end{equation}

\section{Stellar evolution and hydrodynamics codes}
\label{sec:codes}
\subsection{One-dimensional stellar evolution model}
\label{sec:1d-model}
We calculated a 20\,$M_\odot$ stellar evolution model at $Z=0.002$ with an
initial rotation rate $40\%$ of critical velocity
($\upsilon_\mathrm{ini}/\upsilon_\mathrm{crit}=0.4$) with the Geneva stellar
evolution code (GENEC), which is described in detail in
\citet{eggenberger2008a}. The prescriptions for secular instabilities induced
by rotation (secular shear), horizontal shear and
the transport equations (including meridional circulation) used are the same as in \citet{ekstroem2012a},
to which we refer the interested reader. The treatment of dynamical
shear in this model is described in \cref{rcp}. GENEC uses an equation
of state that includes an ideal gas of ions, radiation, and electrons at
arbitrary degeneracy. For the electron part it employs an expansion of the
Fermi--Dirac distribution \citep{kippenhahn1964a,kippenhahn1967a}.

\subsection{Diffusion prescription}\label{rcp}
In the GENEC simulation we adopted the same prescription for dynamical
shear as in \citet{hirschi2004a} using the standard Richardson criterion
as defined in \cref{eq:ri-def}. The critical value, $\Ric=1/4$,
corresponds to the necessary condition for instability in the case of an
initially laminar flow (see derivation in \cref{sec:theory}). It was
therefore used as the limit for the occurrence of dynamical shear in
many stellar evolution codes.

In its full multidimensional nature, shear mixing is not a
purely diffusive process but shows typical features of advection as
well, such as the roll-up of vortex sheets, which temporarily cause a
local inversion of the chemical stratification. If horizontal
averages are considered, shear mixing can be well described by diffusion
\citep[e.g.,][]{prat2014a}. Because of this and due to its ease of
implementation, shear is commonly modeled as such in stellar evolution
calculations. Shear mixing is then added to the other instabilities by
summing the respective diffusion coefficients. Although modeling instabilities
as diffusive processes and summing them up is convenient and thus popular, it
may not be possible to reproduce the effect of certain processes with
diffusion. Furthermore, interactions between the different instabilities might
be more complicated than a simple superposition \citep[e.g.,][]{maeder2013a}.
While a description taking into account the advective nature of shear mixing
would be ideal, the result in \cref{sec:comparison-1d-2d} of this work shows
that the diffusion approximation can go quite far in reproducing a horizontally
averaged hydrodynamics simulation on long timescales.

The diffusion coefficient used for shear mixing in the GENEC
calculations is not of microphysical nature and thus needs a
phenomenological ansatz. In the simulation discussed in this paper we
adopt the diffusion coefficient,
\begin{equation}
D=\frac{1}{3}vl
=\frac{1}{3}\ \frac{v}{l}\ l^2
=\frac{1}{3}\ r\frac{\mathrm{d}\,\Omega}{\mathrm{d}\,r} \ \Delta r^2
=\frac{1}{3}\ r\Delta\Omega\ \Delta r,
\label{formds}
\end{equation}
where $r$ is the mean radius of the zone where the
instability occurs, $\Delta\Omega$ is the variation of $\Omega$ over
this zone and $\Delta r$ is the extent of the zone where $\Ri< \Ric$. It
is derived from the general form of a diffusion coefficient for the
transport of particles with mean velocity~$v$ and mean free path~$l$ as
suggested by J.--P.  Zahn \citep[priv.\ comm., see
also][Sect.~12.2.2]{maeder2009a}.

\Cref{formds} is valid if $\Pe>1$, where $\Pe$, the P\'eclet Number, is the
ratio of advective to diffusive energy transport. Its formal definition
is $Lu/\alpha$, with a typical length scale~$L$, velocity~$u$ and
thermal diffusivity $\alpha$. $\Pe>1$ is fulfilled in the considered
region of the GENEC model. It can be estimated by choosing
the size of the largest eddies of $10^8\,\text{cm}$ as $L$, the
rotational velocity $1.6\times10^8\,\text{cm/s}$ as $u$, and a mean
value of $3\times10^5\,\text{cm$^2$/s}$ as $\alpha$. This gives
$\Pe\approx5\times10^{10}$.
The hydrodynamic simulation was run without diffusion of radiation, which
corresponds to $\Pe \rightarrow \infty$. Neglecting radiative
transport of energy is a reasonable approximation of the physical
situation at the scales of the dynamical shear instability. Due to numerical
diffusion of energy $\Pe$ is limited at a finite value in practice.

\begin{figure}
  \centering
  \includegraphics{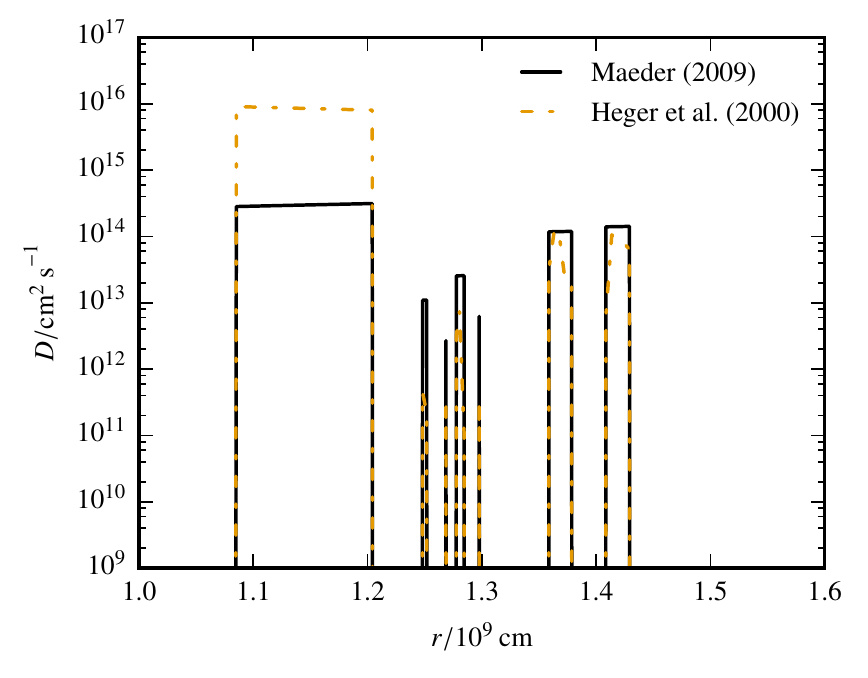}
  \caption{\label{fig:heger-zahn-compare}Comparison of prescriptions for
  the  dynamical shear diffusion coefficients. The underlying stellar
  model is the 20\,$M_\odot$ star described in \cref{sec:1d-model}. The
  black, solid line shows the coefficient defined in \cref{formds}. The
  orange, dash-dotted line was computed using \cref{eq:D-heger}.}
\end{figure}

Other prescriptions for the dynamical shear diffusion coefficient can be
found in the literature.
\citet{heger2000a} use:
\begin{equation}
  \label{eq:D-heger}
  D_\text{Heger}=[\min \{\Delta r,H_P\}(1-\max
\{\frac{\Ri}{\Ric},0\})]^2/\tau_{\text{dyn}}
\end{equation} where
$\tau_{\mathrm{dyn}}=\sqrt{r^3/(GM_r)}$ is the dynamical timescale and
$\Delta r$ the spatial extent of the unstable region.
$M_r$ is the mass enclosed in a sphere of radius $r$.
\Cref{fig:heger-zahn-compare} shows a comparison of this expression
for~$D$ and the one by Zahn for a given stellar model. Both values are very large
($D>10^{12}\,\text{cm}^2\,\text{s}^{-1}$ with many regions showing
$D>10^{14}\,\text{cm}^2\,\text{s}^{-1}$), implying that they cause almost
instantaneous mixing in the affected regions. For comparison, the convective
diffusion
coefficient in the core He burning phase of this star is about
$10^{14}\,\text{cm}^2\,\text{s}^{-1}$. In practice, it is much more important which zones
are flagged as being unstable to dynamical shear than what the exact value of
$D$ is. Both prescriptions agree in this respect.

\citet{brueggen2001a} studied the dependence of $D$ on $Ri$ in 3D
hydrodynamic simulations of plane-parallel flows. As already
mentioned in \cref{sec:intro-motivation}, their use of the Boussinesq
approximation for the definition of \Ri{} makes it impossible to
reconstruct the dependence on the full definition of \Ri{} from the
given data.

\begin{figure*}
  \centering
  \includegraphics{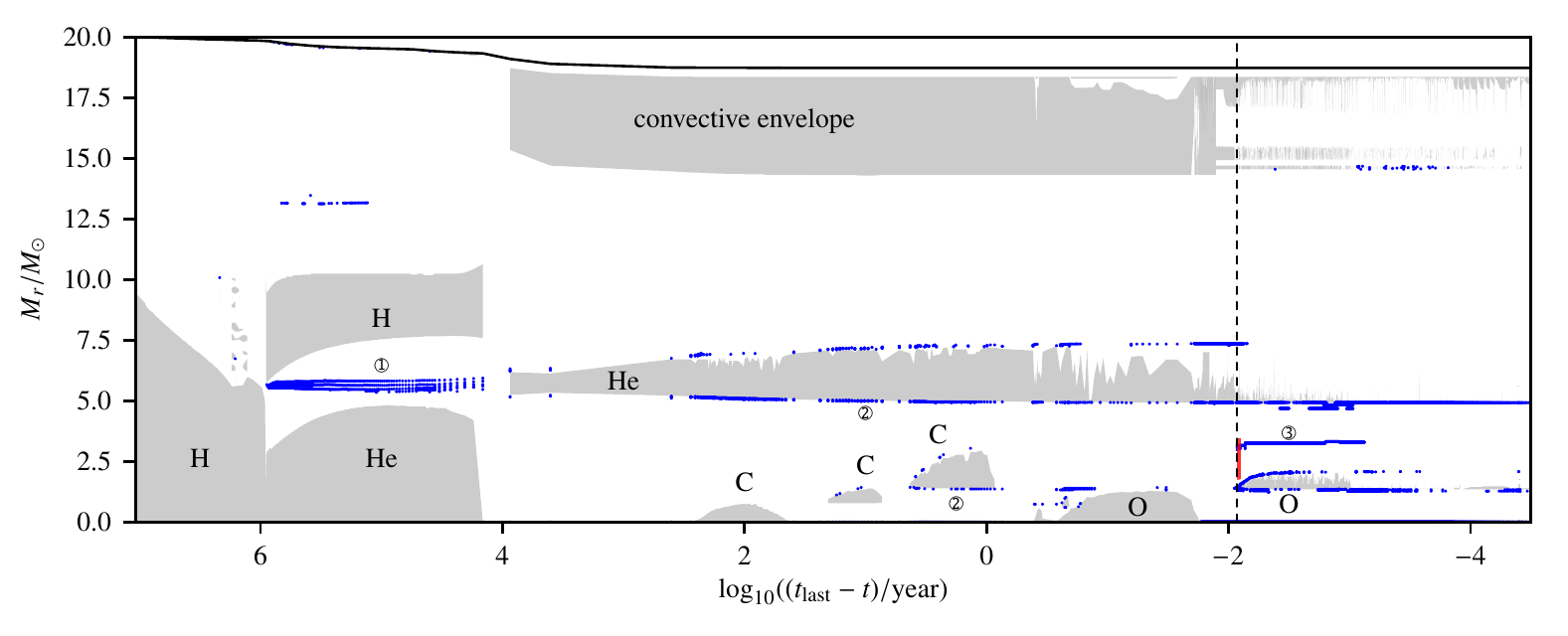}
  \caption[Structure evolution
  diagram]{\label{P020z02S407_full}Structure evolution diagram of the
  $20\,\text{M}_\odot$ star with $Z=0.002$ and an initial rotation rate
  $40\%$ of critical velocity ($\upsilon_{\rm ini}/\upsilon_{\rm
  crit}=0.4$).  The ordinate represents the enclosed mass and the
  abscissa the time left until the last model, which is in-between core
  oxygen burning and core silicon burning. The gray zones indicate
  convective zones. The blue dots mark shells which are
  unstable to dynamical shear. They can either mark a single
  unstable zone or an extended region, which is bounded by shear stable
  layers. The circled numbers refer to different situations in which
  dynamical shear occurs in stellar models as given in the list in
  \cref{rcp}. The core and shell burning regions are indicated with the
  corresponding element symbol. The vertical dashed line indicates the time
  that was used as the initial condition for the 2D~hydrodynamic simulation.
  The red area marks the spatial and temporal extent of the 2D
  simulation.}
\end{figure*}

\begin{figure}
  \centering
  \includegraphics[width=0.49\textwidth]{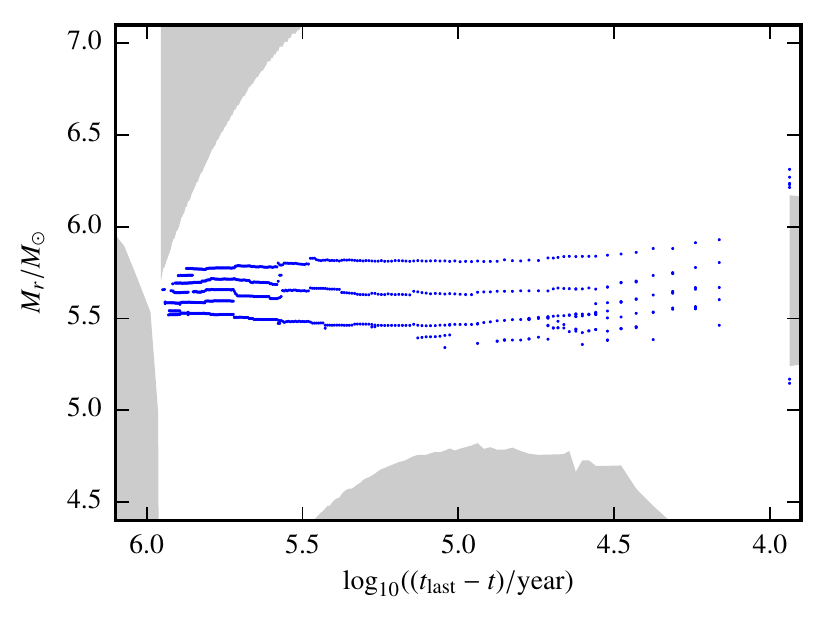}
  \caption{Same as
  \Cref{P020z02S407_full} but zoomed in on unstable region at the
  edge of what was the hydrogen-burning convective core before
  the beginning of helium burning. The blue dots indicate zones which
  are unstable to dynamical shear. Each dot indicates an
  extended shear unstable zone, which is surrounded by shear stable
  layers. The gray areas are convective regions.
  \label{P020z02S407_he-shell}}
\end{figure}

\Cref{P020z02S407_full} shows the structure evolution diagram for the
20\,$M_\odot$ model, where convective zones are shaded in gray. The blue
dots indicate shells which are unstable to dynamical shear.
These can be single zones or extended regions bounded by shear
stable layers. This figure thus shows three main types of zones that are
unstable to dynamical shear.

The first type appears in the radiative layers from the end of the main
sequence onwards. The unstable zones are the result of the combined general
core contraction and envelope expansion at the end of core hydrogen burning.
One major unstable region appears around the mass coordinate $M_r \approx
6\,M_\odot$, which corresponds to the edge of the former hydrogen-burning
convective core. It is caused by core contraction. It is marked by
\textcircled{\raisebox{-0.8pt}{\texttt{1}}} in \cref{P020z02S407_full}. A zoom
in on this purely radiative, shear unstable region is shown in
\cref{P020z02S407_he-shell}. At the same evolutionary stage envelope expansion
causes another small unstable region between the hydrogen-burning convective
shell and the envelope of the star.

The second type is present at convective zone boundaries, where convection
redistributes angular momentum very efficiently. This creates a sharp angular
velocity gradient at convective zone boundaries, which lowers \Ri{} and is
therefore responsible for development of dynamical shear. Examples are marked
by \textcircled{\raisebox{-0.8pt}{\texttt{2}}} in \cref{P020z02S407_full}.

The third type occurs at the interface of  carbon- and neon-rich shells after
core oxygen burning. Following the end of core oxygen burning, most of the star
undergoes a contraction until the oxygen-burning convective shell develops.
This contraction increases the angular velocity gradients and thereby causes
the top of the last convective carbon-burning shell to become unstable to
shear. It is indicated by \textcircled{\raisebox{-0.8pt}{\texttt{3}}} in
\cref{P020z02S407_full}.

In this first study of dynamical shear in massive stars, we decided to focus on
the last point, unstable zone during the advanced phases, for the following
reasons. The evolutionary time scale after core oxygen burning allowed us to
cover a significant part of the 1D~stellar evolution model in
2D~hydrodynamics. In particular we could directly compare the behavior of the
instability in both dimensionalities. The smaller length scales of the
carbon and neon shells made the computation easier than, for example, the
helium shell.

  \subsection{The Seven-League Hydro code}
  \label{sec:intro-slh}
  The numerical tool for the hydrodynamic simulation of the
  dynamical shear instabilities was the Seven-League Hydro (SLH) code.
  SLH solves the compressible Euler equations with a finite-volume
  scheme in one, two, or three spatial dimensions. It offers a
  range of numerical flux functions, which are specifically tuned to
  yield accurate results in all Mach number regimes. The calculations
  presented in this article are computed with the Roe--Miczek method
  \citep{miczek2015a}. SLH offers the choice of implicit or explicit
  time stepping for the solution of the hydrodynamics equations.
  Implicit time stepping is more efficient in the low Mach number regime
  ($M\lesssim 10^{-2}$). As the maximum Mach number in the inertial
  frame is 0.5, we choose explicit time integration for the simulation
  presented in this paper. This cannot be fixed by using a rotating frame of
  reference because the spread in Mach number between the inner- and outermost
  layers amounts to 0.1. The explicit time stepping method of choice in SLH is
  RK3 \citep{shu1988a}, which is second-order accurate and has excellent
  stability properties.

  A number of different grid geometries are implemented in SLH with the
  mapped-grid formalism \citep[e.g.,][]{calhoun2008a,kifonidis2012a}.
  This allows us to use an arbitrary, structured, curvilinear grid,
  which is adapted to the problem geometry, but performing all the
  computations on a logically rectangular grid. The connection between
  this computational grid and the curvilinear grid is established
  through a mapping function. Its derivatives enter the finite-volume
  discretization at several places. In the present case we employed this
  quite general formalism to create a two-dimensional polar grid as it
  is best adapted to the geometry of the rotating problem.

  The equation of state of choice in this context was the Helm\-holtz
  equation of state \citep{timmes2000a}, which includes the
  contributions of radiation, an ideal gas of the nuclei
  and a Fermi gas of electrons at arbitrary degeneracy. We did not switch
  on Coulomb corrections. This choice was due to the fact that the
  equation of state in GENEC -- on which our initial conditions
  are based -- does not include them. In the considered range of the
  star their contribution to pressure and energy is less than 1\%. The
  treatment of electrons in the partially degenerate regime is
  significantly different in the two codes. To retain the same regions of
  convective stability in stellar evolution and hydrodynamics we needed to
  employ a special mapping procedure detailed in \cref{sec:model_setup}.

\section{Model setup}\label{sec:model_setup}
  Mapping a one-dimensional stellar evolution model to a
  multidimensional hydrodynamics code can introduce subtle problems,
  even if it does not involve rotation. The challenge is to keep the
  model in hydrostatic equilibrium and preserve many of the
  thermodynamic properties well, while interpolating on an often much
  finer grid and possibly applying some smoothing on the very steep
  gradients introduced by the prescription for convection in the
  stellar evolution model. Another source of error is the fact that the
  stellar evolution code might use a different equation of state than
  the hydrodynamics code.

  The condition of hydrostatic equilibrium for a cylindrically symmetric,
  rotating star is \citep[e.g.,][]{maeder2009a}
  \begin{align}
    \label{eq:hystat}
    \PD{P}{r} - \rho \left(g + r \sin^2 \vartheta \Omega^2\right) &= 0,\\
    \label{eq:hystat-theta}
    \PD{P}{\vartheta} - r \sin \vartheta \cos \vartheta \Omega^2 \rho &= 0.
  \end{align}
  This assumes the Roche model, meaning that gravity is supposed to be the
  same as if the mass enclosed in the isobaric shell were located at the
  center of the star. This implies that the gravitational acceleration
  is always directed to the center of the star.

  A natural choice for performing 2D~simulations of a rotating star is
  the equatorial plane. This is why the following considerations are
  done in the equatorial plane (colatitude $\vartheta = \pi / 2$) of the
  star only. The $\vartheta$ derivative in \cref{eq:hystat-theta}
  vanishes in this case. To compute the radial pressure profile we can
  rewrite \cref{eq:hystat} as an ordinary differential equation~(ODE)
  for pressure with radius~$r$ as the independent variable. We note that,
  in contrast to \cref{eq:bvf}, $g(r)$ carries a sign, as the vector
  $\vec{g}$ points in negative $r$-direction. This leads to
  \begin{equation}
    \label{eq:hystat-ode}
    P'(r) = \left(g(r) + r \Omega(r)^2\right) \rho(r)
          \equiv g_\text{eff}(r) \rho(r).
  \end{equation}
 In this form $\rho(r)$ is just the (interpolated) density profile of
 the stellar model. An alternate way of computing $\rho$ is to take the
 temperature profile of the model and use the equation of state to get
 $\rho(T(r), P(r))$. This is needed for reproducing temperature
 accurately. The same could be done for entropy by using $s(r)$ and
 computing $\rho$ as a function of $s(r)$ and $P(r)$ from the equation
 of state. Here, $P(r)$ is not the value taken from the stellar
 model but the dependent function in the differential equation. What
 makes a good choice for the density profile depends on what needs to be
 reproduced accurately for the problem.

\begin{figure}
    \includegraphics{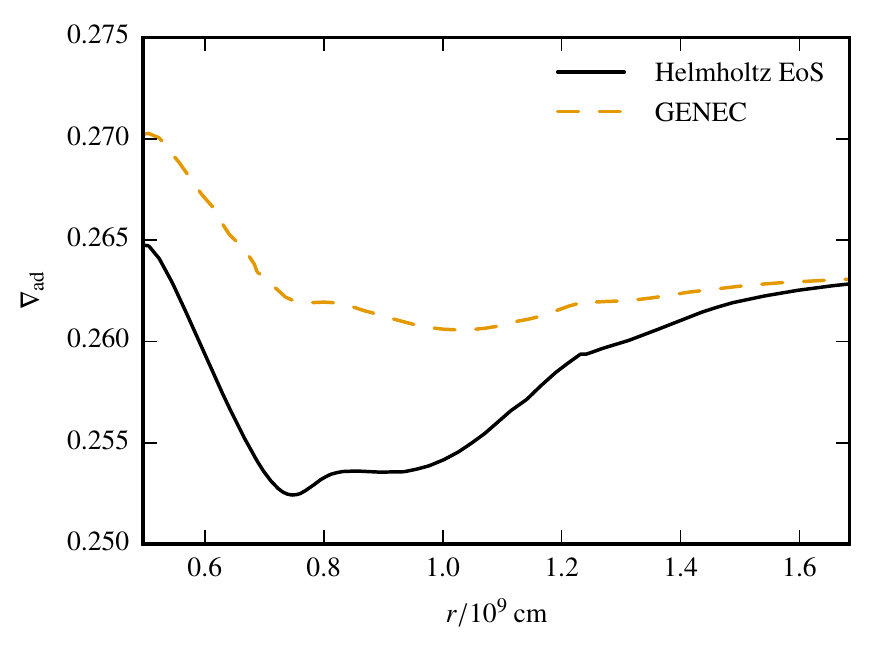}
    \caption{\label{fig:nabad-genec-helm}
    $\nabad$ computed from the equation of state as it is used in GENEC
    (dashed line) and the value obtained from the Helmholtz equation of
    state using the same values for density, temperature, and
    composition as input.}
  \end{figure}

\begin{figure}
    \includegraphics{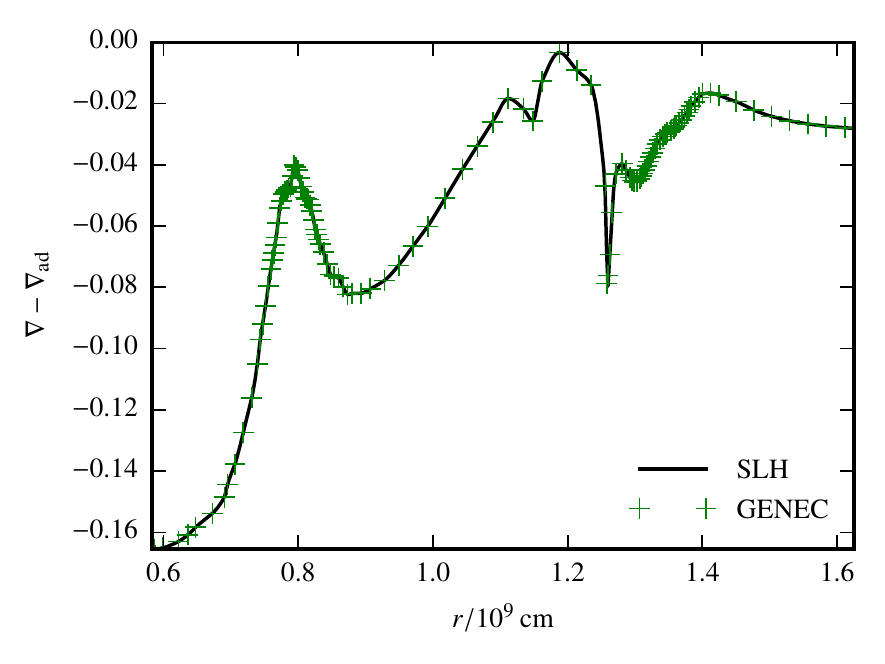}
    \caption{\label{fig:nabla-nabad} Comparison of the quantity
    $\nabla - \nabad$ from the GENEC model with the reconstructed values for
    the Helmholtz equation of state using \cref{eq:hystat-ode-nabla}. This
    ensures that radiative regions stay radiative after mapping the original data
    to SLH.}
  \end{figure}

\begin{figure}
    \includegraphics{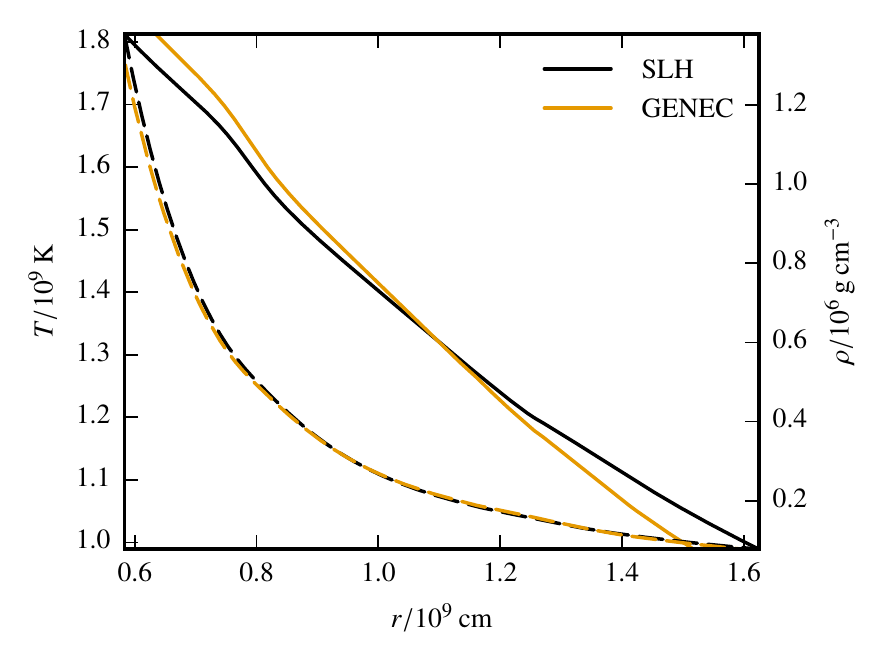}
    \caption{\label{fig:rhoT-reconst}Radial profiles of temperature (solid
    lines) and density (dashed lines). The yellow lines are the original
    data from GENEC and the black lines are the initial data for the
    simulation with SLH, which are reconstructed using the method
    described in \cref{sec:model_setup}.}
  \end{figure}

  In the case of the current model the most pressing concern was the
  difference in the equation of state. SLH uses the Helmholtz equation
  of state while GENEC employs a simpler approximation to a partially
  degenerate electron gas (see \cref{sec:intro-slh}). While the
  relative difference in pressure is less than $10^{-3}$
  \citep{kippenhahn1964a}, there is a significant discrepancy in the
  value of $\nabad$, as shown in \cref{fig:nabad-genec-helm}. It is
  clear from the definition of the \BVF that reducing $\nabad$
  even by a small amount can cause marginally stable regions in the star
  to become convectively unstable. As the model was specifically chosen
  to be convectively stable, this effect is problematic. This is why we
  used input quantities that differ from the typical choice of density or
  temperature. Using entropy was not easily possible as it is not readily
  available from the stellar evolution code. Instead the new model was
  constructed to reproduce the quantity $\nabla - \nabad$, which
  determines convective stability, exactly. To this end we needed to
  extend the ODE \cref{eq:hystat-ode} to a system for pressure~$P$ and
  temperature~$T$. Here, Density~$\rho(P,T)$ is computed as a function of
  these two quantities. An expression for $T$ can be derived from
  \cref{eq:nabla},

  \begin{equation}
    \nabla = \D{\ln T}{\ln P} = \frac{P}{T} \D{T}{P}
           = \frac{P}{T} \D{T}{r} \D{r}{P} = \frac{P}{T\rho
           g_\text{eff}} \D{T}{r}.
  \end{equation}
  To keep $\nabla - \nabad$ identical in the stellar evolution model and
  the initial setup of the 2D~hydrodynamic simulation, we did not use
  $\nabla(r)$ as an input quantity directly but instead we defined
  \begin{equation}
    \nabla = \left(\nabla - \nabad\right)_\text{model} + \nabad(P,T),
  \end{equation}
  where the expression in the parentheses with subscript ``model'' was
  taken from the stellar evolution model and $\nabad(P,T)$ was given by
  the Helmholtz equation of state, employed in the hydrodynamics code.
  This leads to a new coupled system of ODEs,
  \begin{equation}
    \label{eq:hystat-ode-nabla}
    \begin{aligned}
      P'(r) &= g_\text{eff}(r) \rho(P,T),\\
      T'(r) &= \frac{T\rho(P,T) g_\text{eff}(r)}{P}
               \left[ \left(\nabla - \nabad\right)_\text{model} + \nabad(P,T) \right].
    \end{aligned}
  \end{equation}
  The solution of this system matches the convective regions of the
  stellar evolution code by construction as illustrated in
  \cref{fig:nabla-nabad}. This even works in the case of a different
  equation of state. A drawback of this approach is that all other
  thermodynamic quantities are not guaranteed to reproduce the original
  model. Fortunately, the deviation is minor for the case at hand as seen in
  \cref{fig:rhoT-reconst}. The condition for the development of a dynamical
  shear instability is not changed by this mapping procedure as $\Omega$ and
  $N^2$ are identical to the original stellar model.

  This approach does not ensure gravity to be consistent with the density
  field. This could easily be fixed by extending \cref{eq:hystat-ode-nabla}
  with an additional equation for the enclosed mass~$m$. We did not follow this
  approach here because the simulation is carried out with a fixed, external
  gravity field anyway.

  The values for the radius that are given in the GENEC simulation are
  averages over isobaric shells. They correspond to the value at
  $\vartheta = \arccos \sqrt{1/3} \approx 54.7^\circ$, the root of the
  Legendre polynomial $P_2(\cos \vartheta)$. The shape of the
  isobar~$r(\vartheta)$ can be computed from the profiles of $\Omega$.
  We used the computed value of the equatorial radius and imposed the
  average $\nabla - \nabad$ on the isobar there. While this is clearly
  not an exact reconstruction of shellular rotation, as $\nabla -
  \nabad$ is not constant on isobars, it yields a 2D~model that is
  qualitatively similar to the 1D~code, meaning that it has the same convective
  zones and $\Omega$ profile. The relative change in $T$ and $\rho$
  is less than 8\%.  An exact reconstruction of the original profile
  that also preserves convective regions is not possible in this case
  due to the differences in the equation of state.

\begin{figure}
    \includegraphics{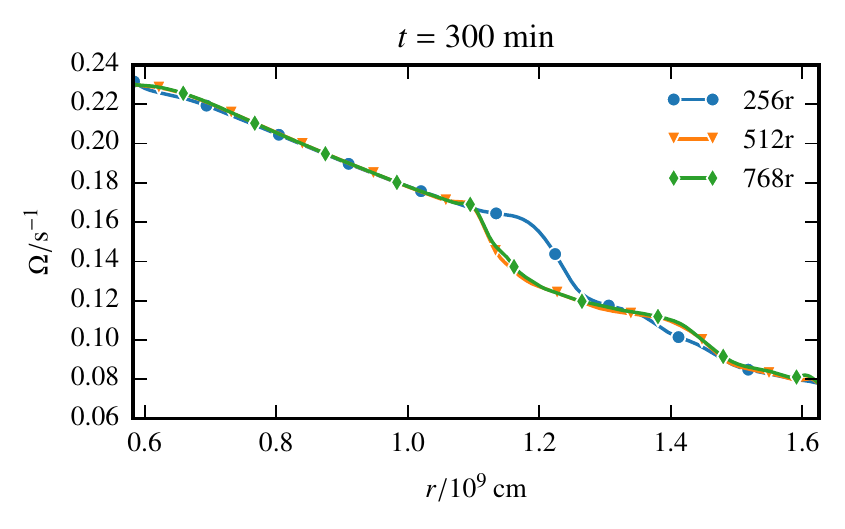}
    \caption{\label{fig:res-omega}Angular average of angular velocity at
    $t=300\,\text{min}$ after the shear mixing has stopped. The lines
    show the profile for simulations with 256, 512, and 768 radial
    zones. Each have the corresponding number of angular zones to keep
    the aspect ratio constant.}
  \end{figure}

  For the radial boundary we used an impenetrable wall boundary on the
  inner side. This type of boundary condition does not employ
  ghost cells but directly sets the flux at the interface. The details
  can be found in \citet{ghidaglia2005a}.  At the outer boundary we left
  the ghost cells at their initial value.  This allows in- and outflow
  according to the internal flow, while preventing runaway effects
  caused by extrapolation of the adjacent velocities.  There was no
  boundary in azimuthal direction as we cover the full $2\pi$ azimuthal
  angle ($\phi$) of the equatorial plane and the geometry is
  intrinsically periodic. The grid resolution was $512(r)\times
  1024(\phi)$ with equidistant spacing. As a convergence test we
  computed a higher resolution run with $758(r)\times1536(\phi)$ grid
  cells and a lower resolution run with $256(r)\times512(\phi)$. The
  final angular momentum profile after mixing has stopped is shown in
  \cref{fig:res-omega}. It demonstrates that the instability starts in all
  three cases but convergence is only reached with at least 512 radial zones.
  A study of the second-order structure function of kinetic
  energy at $t=50\,\text{min}$ showed that the characteristic length scale of
  the grids with 512 and 758 zones is $1\times10^8\,\text{cm}$, while it is $3
  \times 10^8\,\text{cm}$ in the case of 256 zones. The analysis in this study
  was done for the simulation with 512 radial zones. At this resolution
  the average hydrodynamic time step is 1.3~ms. For comparison, the typical GENEC
  time step at this stage of evolution is about 5~s.

  We did not include any source terms except gravity. Both nuclear
  reactions and plasma neutrino losses have an influence on the
  evolution of the instability as well. This is, for example, through
  the mixing of unburned carbon into the hotter regions below and the
  resulting increased energy release. The minimum time scale of
  plasma neutrino losses on the simulation domain, given by the ratio of
  the total energy to the loss rate, is roughly 300~h. It was estimated
  using the fitting formulae of \citet{itoh1996a}. Nuclear burning
  acts on the same time scale, although in a different region
  of the domain. The energy release rate was calculated using one-zone
  calculations with the nuclear network YANN \citep{pakmor2012b}.
  Comparing this to the duration of the simulation of $\sim$6~h suggests
  that both effects are relevant in the dynamic evolution. In our
  present study, however, we focused purely on the hydrodynamic effects.
  This was done intentionally to disentangle the effect of the
  source terms and the hydrodynamic development of the instability, even
  tough it impedes the possibility to compare the hydrodynamic
  simulation and the stellar evolution model directly because the
  latter includes the aforementioned source terms.

\section{2D~hydrodynamic simulation of dynamical shear}\label{sec:hydro}
\begin{figure*}
    \centering
    \includegraphics{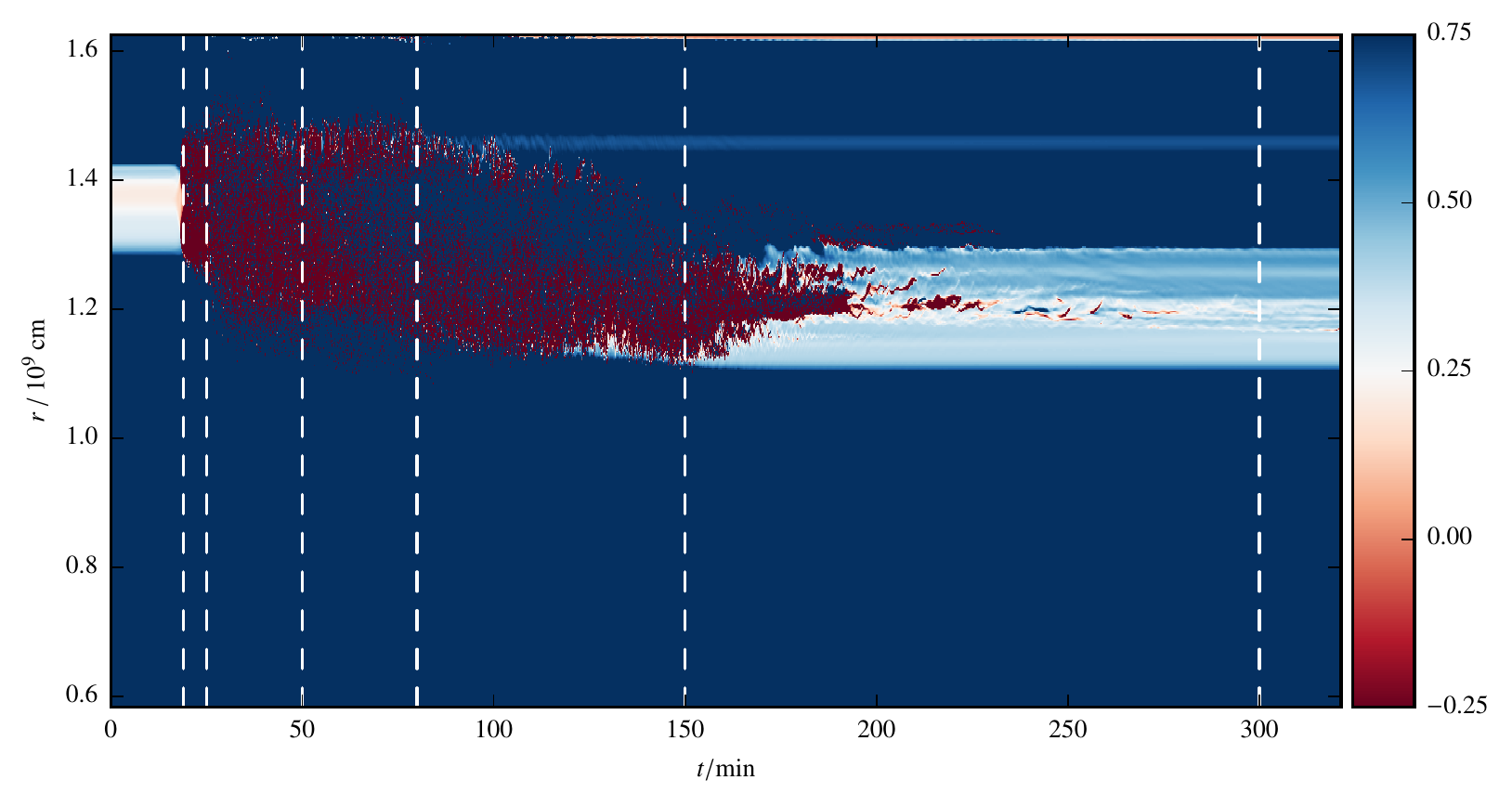}
    \caption{\label{fig:ri-contour}Time evolution of the averaged
    Richardson number. Blue regions are expected to be stable to
    dynamical shear. Red regions are Richardson unstable. The Richardson
    number was computed locally on the 2D data and averaged
    afterwards. The vertical dashed lines correspond to the 2D~snapshots
    shown in \cref{fig:ri-panels,fig:abar-panels}.}
  \end{figure*}

  Our goal was to study the development of
  dynamical shear instabilities in a realistic stellar environment. This
  was achieved by transferring a model from the stellar evolution code
  GENEC to the hydrodynamic code SLH\@.  To get an overview of the
  temporal evolution of the shear instability it is useful to visualize
  the evolution of the Richardson number, \cref{eq:ri-def}, in different
  parts of the simulation domain.  We computed the Richardson numbers
  locally on the 2D~data and averaged it then on
  each radius instead of obtaining a 1D~profile of the other quantities
  and calculating \Ri{} from that. This ordering of the averaging
  operation makes regions containing instabilities more obvious as
  radial bins containing only few unstable zones are dominated by the
  often strongly negative values of \Ri{} there. \Cref{fig:ri-contour}
  shows the temporal evolution of angular averages of \Ri. This
  highlights the region between $1.3\times 10^9$ and $1.4\times
  10^{9}$~cm, in which we expect the initial development of the shear
  instability from the well-known theory outlined in \cref{sec:theory}.
  All figures show the simulation with a radial resolution of 512 grid
  cells.

\begin{figure*}
    \includegraphics{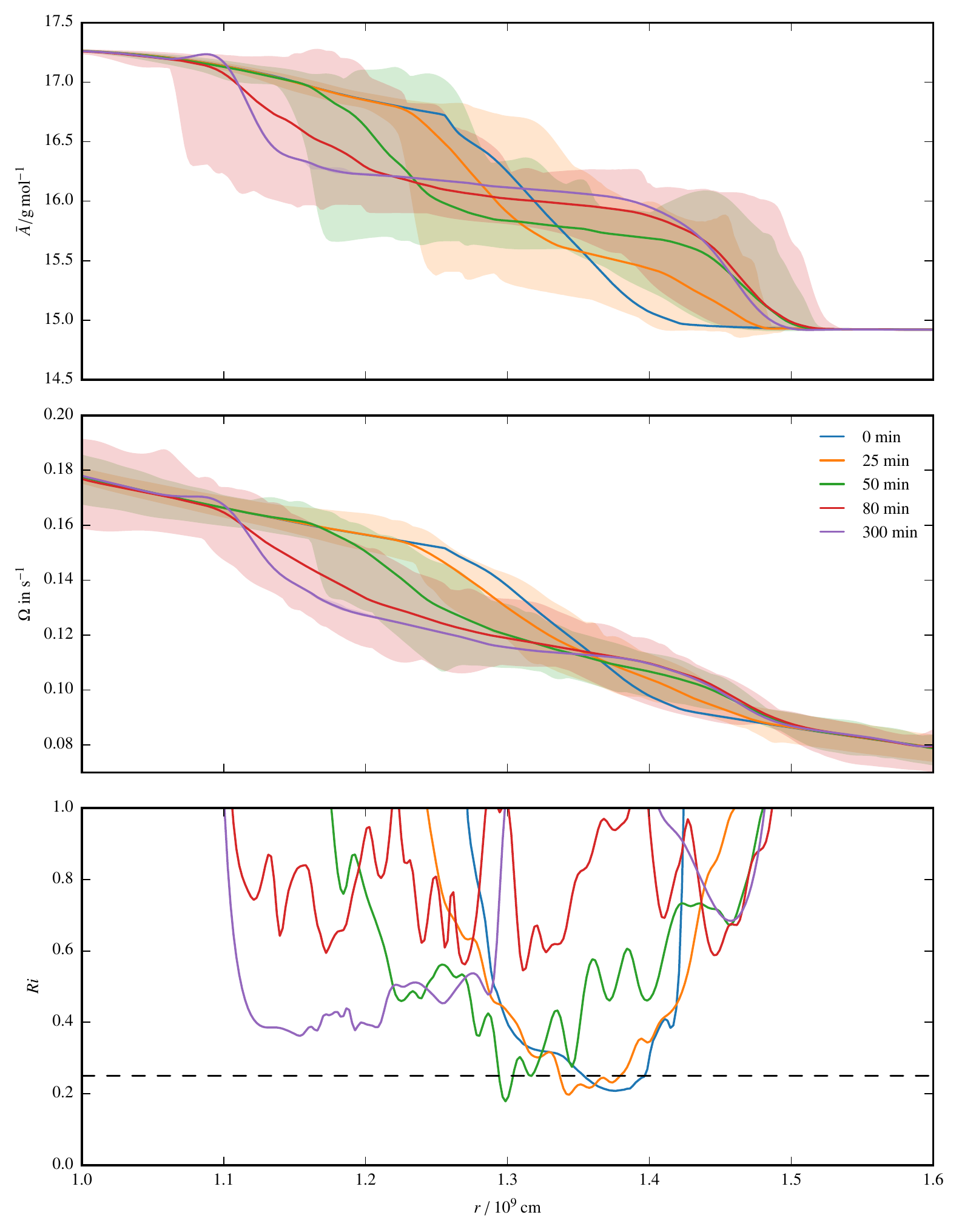}
    \centering
    \caption{\label{fig:profile-lines}Angular average of mean atomic
    mass~$\bar{A}$, angular
    velocity~$\Omega$, and Richardson number~\Ri{} for snapshots at
    different times indicated by line color. The regions shaded in the
    respective line color show the range between the minimum and maximum
    value at the respective radius.}
  \end{figure*}

\begin{figure*}
    \centering
    \includegraphics{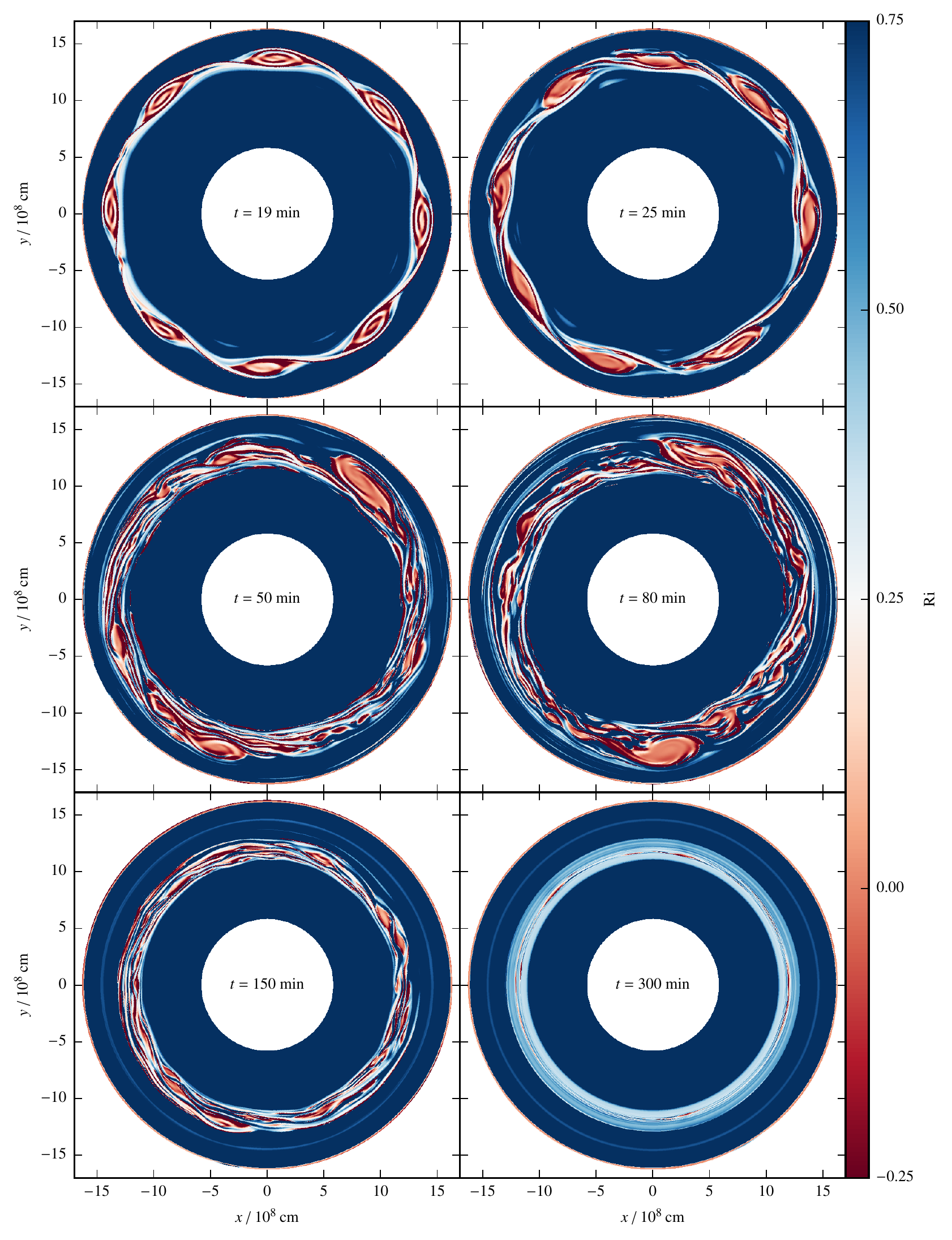}
    \caption{\label{fig:ri-panels}Richardson number in the equatorial
    plane. The panels correspond to the time steps indicated by vertical
    lines in \cref{fig:ri-contour}. \Ri{} is evaluated along radial
    profiles. The blue regions are are stable to shear, the red regions are
    unstable. A movie of the entire simulation is available at
    \url{https://slh-code.org/papers/dynshear}.}
  \end{figure*}

  The first obvious result is that the instability developed at the
  position predicted by the Richardson criterion (indicated by the
  blue line in the bottom panel of \cref{fig:profile-lines}, which is
  the initial Ri profile in the 2D simulation). We see the
  growth of the instability to a noticeable level after 19~min of
  simulation time. The spatial distribution of \Ri{} at this time is
  depicted in the first panel of \cref{fig:ri-panels}. It is still
  fairly symmetric, dominated by one large-scale mode. This regular
  pattern is caused by the radially symmetric initial conditions. It
  shows the typical rolled-up vortex sheets of a Kelvin--Helmholtz instability.

\begin{figure*}
    \centering
    \includegraphics{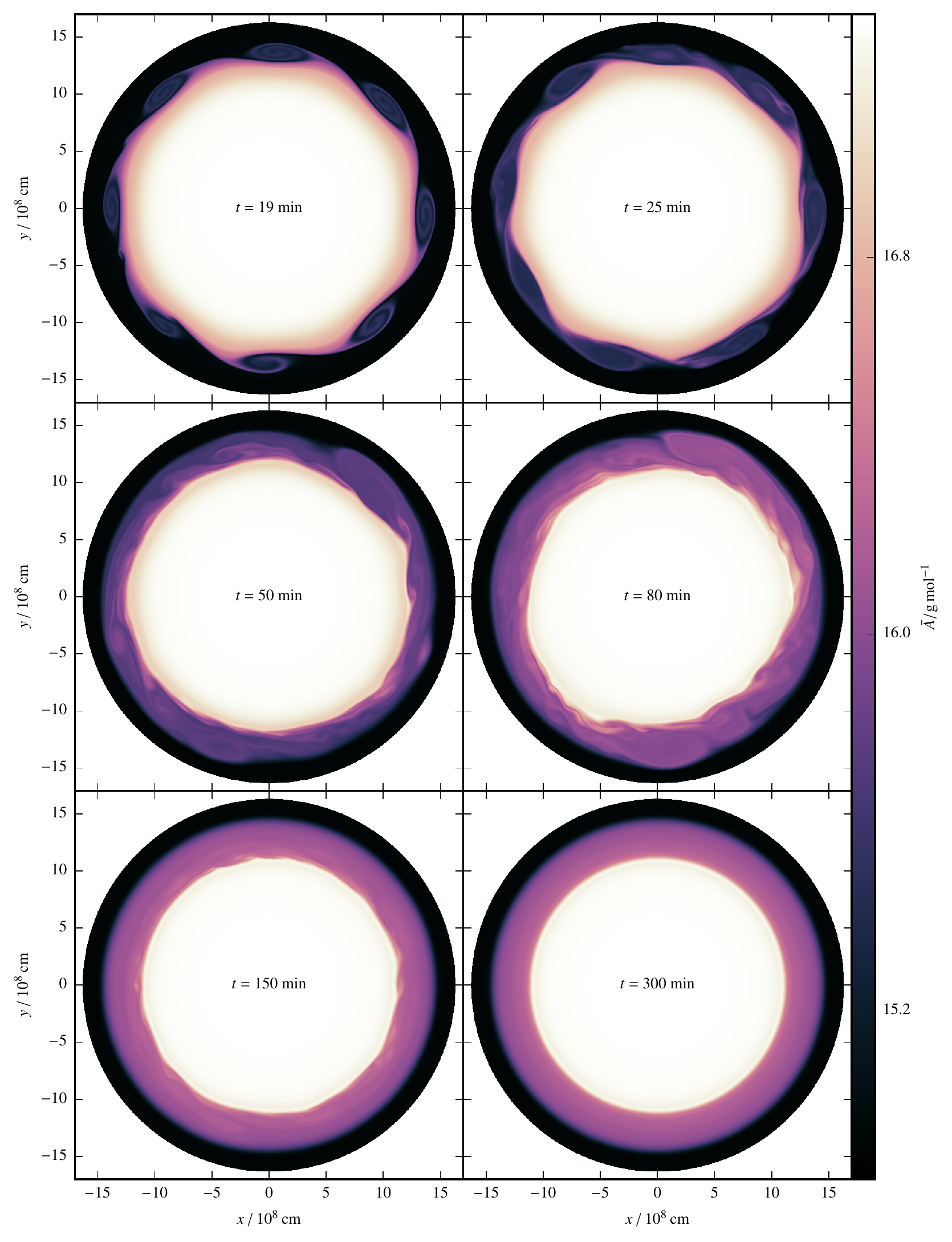}
    \caption{\label{fig:abar-panels}Mean atomic mass~$\bar{A}$ in the
    equatorial plane. The panels correspond to the time steps indicated
    by vertical lines in \cref{fig:ri-contour}.}
  \end{figure*}

  The next panel in \cref{fig:ri-panels,fig:abar-panels} shows the
  departure from this very regular pattern at $t=25\,\text{min}$. In
  the following panels at $t=50\,\text{min}$ and $t=80\,\text{min}$ the
  flow develops a chaotic structure. This creates a mixed region
  surrounded by stable layers with the original stratification on top
  and bottom. The result is a smooth step profile
  in $\bar{A}$ and angular velocity~$\Omega$. The lower edge of this
  step is the position at which new, smaller shear instabilities
  develop, causing a further change in the $\Omega$ profile. One case of
  this is seen at $t=150\,\text{min}$.  As a consequence the region
  affected by the instability expands to lower radii as could already be
  seen in \cref{fig:ri-contour}.

  \Cref{fig:ri-contour} indicates that the unstable regions become
  smaller after $t=150\,\text{min}$ and have completely disappeared by
  $t=300\,\text{min}$. Shear mixing has brought $\Omega$ back into the
  \Ri-stable regime. The $\bar{A}$ and $\Omega$ profiles go back to a
  almost perfect cylindrical symmetry. This is illustrated in the first
  two panels of \cref{fig:profile-lines}, where the angular means and
  the minimum and maximum values almost coincide for
  $t=300\,\text{min}$.

  The bottom panel of \cref{fig:profile-lines} shows the development of
  the radially averaged profile of the Richardson number. We see how the
  region with $\Ri<0.25$ moves from its initial place between 1.35 and
  $1.4\times 10^9\,\text{cm}$ to smaller radii. In the final state at
  $t=300\,\text{min}$ there is a significant region where
  $\Ri\approx0.4$. This indicates that $\Ri<0.25$ is indeed the right
  criterion for the development of dynamical shear instabilities in this
  case.

\section{Comparison between 1D and 2D~models}
\label{sec:comparison-1d-2d}
\begin{figure*}
  \centering
  \includegraphics{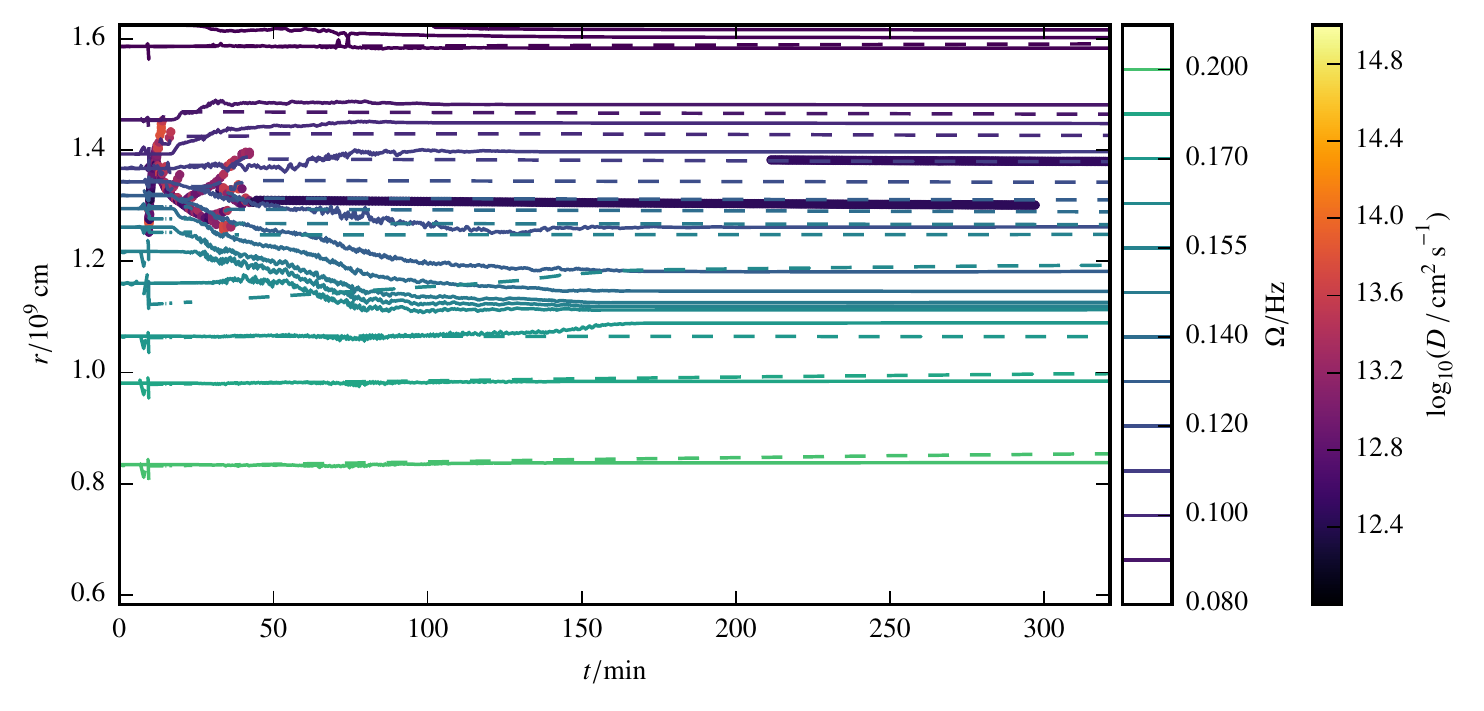}
  \caption{\label{fig:omega-contour}Contour lines of the time evolution
  of the angular velocity~$\Omega$ profile. The solid lines show radial
  averages of the 2D~SLH simulation. The dashed lines show the evolution
  in the 1D~GENEC model with the radial coordinate projected into the
  equatorial plane. The dots show the location of shear unstable regions
  in the GENEC simulation. The color of the dots indicate the magnitude
  of the GENEC diffusion coefficient.}
\end{figure*}

The exact time of the onset of the instability in the SLH simulation is
marginally later than in the GENEC model as seen in
\cref{fig:omega-contour} but this delay has little significance as it mostly
depends on the exact time step used for mapping from 1D to 2D. The
hydrodynamic simulation starts from a laminar flow so the shear
instability needs to grow from an initial perturbation at the level of
numerical noise.

In both the 1D and the 2D~model the instability affects the $\Omega$
profile for about 150~min (see \cref{fig:omega-contour}), after which it
stays virtually constant. The GENEC model still has regions in which
dynamical shear is active but with a low value of~$D$, so the effect
cannot be seen on the timescales considered here. In contrast to the
1D~model, the 2D~simulation will not change at later times as we do not
include any evolutionary processes like nuclear reactions, neutrino
losses, contraction, or expansion of the stellar core.

\begin{figure*}
  \centering
  \includegraphics{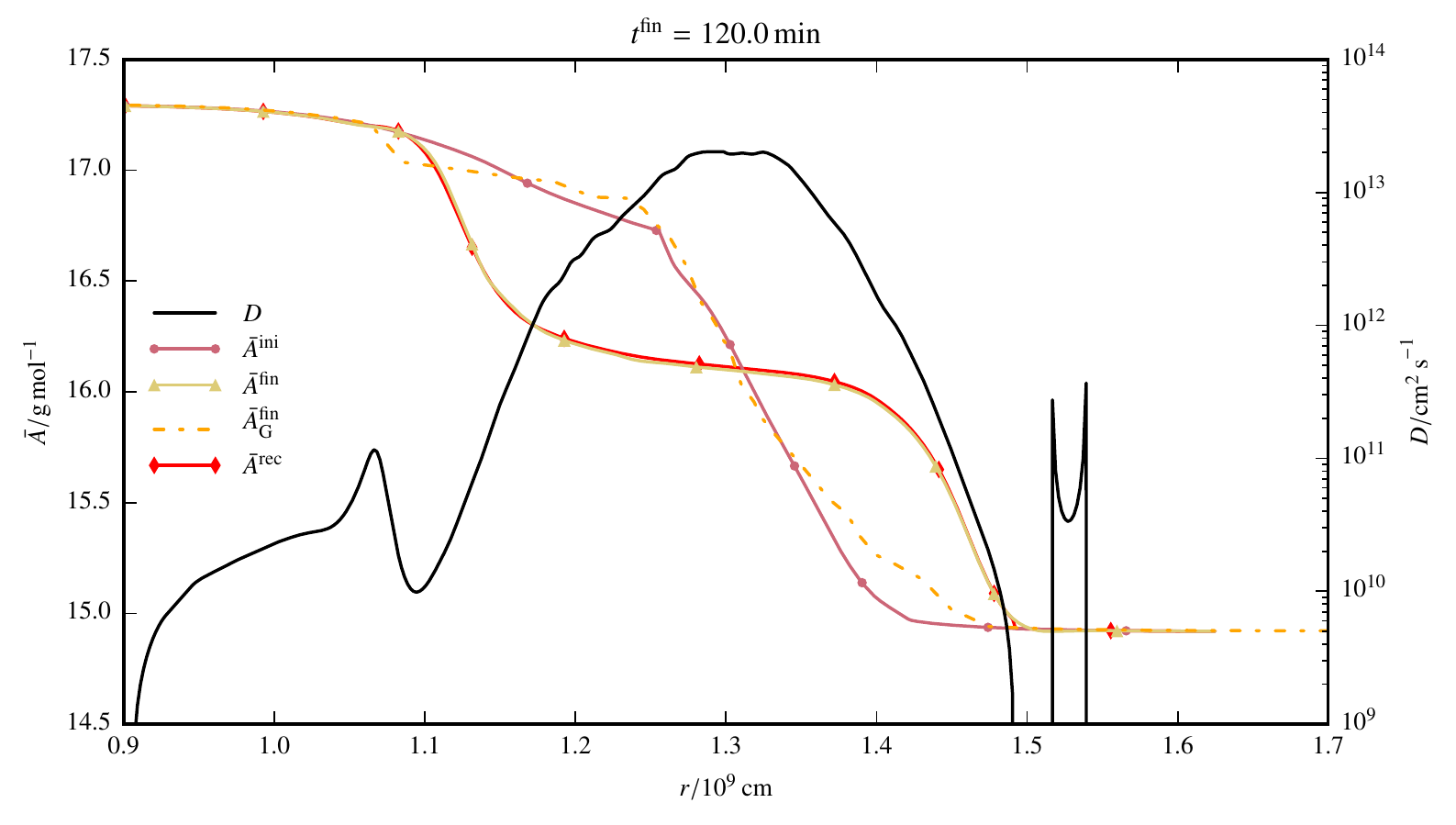}
  \caption{\label{fig:Dshear-2D}Reconstructed diffusion coefficient from
  initial and final profiles of mean atomic mass. The initial profile
  of~$\bar{A}^\text{ini}$, the final profile of the hydro
  simulation~$\bar{A}^\text{fin}$, and the corresponding final profile
  from GENEC~$\bar{A}_\text{G}^\text{fin}$ are shown. The final state is
  taken at $t=120\,\text{min}$. The solid black lines shows the
  effective diffusion coefficient~$D$ from the hydro simulation.
  Negative values due to locally antidiffusive behavior are set to 0.
  The result of $D$ being applied to the initial $\bar{A}$~profile is
  shown as~$\bar{A}^\text{rec}$.}
\end{figure*}

\Cref{fig:omega-contour} gives an overview of the rearrangement of the
angular velocity~$\Omega$ profile in both the 1D~hydrostatic and
2D~hydrodynamic simulation. We see that the instability occurs in both
cases shortly after the start of the simulation. Another similarity is
that the original instability is quenched by the change in $\Omega$ and
new unstable regions are formed in neighboring layers. This is where the
details begin to differ. In the 1D~case the initial instability evolves
much faster. This is due to the fact that a very high diffusion
coefficient ($D>10^{13}\,\mathrm{cm}^2\,\mathrm{s}^{-1}$) is used in the
unstable layer and a very low value in neighboring zones. This moves the
strong $\Omega$ gradient to a neighboring cell rather than smoothing it
as in the 2D~model. After about 50~min the major changes are over and a
one-zone, \Ri~unstable region remains for a time much longer than in the
hydro simulation. This does not have a significant impact on the
long-term evolution of the $\Omega$ profile as the diffusion coefficient
is rather weak in the used prescription. The plot of $\bar{A}$ in
\cref{fig:Dshear-2D} shows that the region affected by shear mixing agrees
between 1D and 2D but the magnitude is strongly underestimated by the
1D~models. The latter only deviate slightly from the initial condition, whereas
the 2D case shows a smooth step profile. If a similar behavior were
realized in the GENEC model, this would likely have a noticeable impact on its
later evolution. While 3D~hydrodynamic simulations are needed to make final
quantitative statements, these results indicate that the effect of dynamical
shear might be underestimated in its current treatment in stellar evolution
codes.

In 1D~models mixing due to dynamical shear was implemented by solving a
diffusion equation with a temporally and spatially varying diffusion
coefficient (see \cref{rcp}). In the framework of 2D or 3D~hydrodynamics such a
coefficient is not readily available as the process is not fundamentally
of diffusive nature. To ease the comparison the two kinds of simulation
it is useful to know what effective diffusion coefficient would be
needed to mimic the effect of hydrodynamics, at least on the horizontal
averages. We followed the approach of \citet{jones2017a} to compute the
diffusion coefficient~$D$ necessary to transform the initial profile of
$\bar{A}$ to the final profile in one diffusion step. This assumes a constant
$D$ over 120~min. The diffusion step is much longer than the time steps of the
stellar evolution calculation in this phase and the result is therefore not directly
suited for replacing the prescriptions mentioned in \cref{rcp}. Still, it gave
us a measure of the net diffusion caused by the shear instability. The initial
and final profiles were obtained by averaging over several hydrodynamics
snapshots, corresponding to one GENEC time step.

\begin{figure*}
  \centering
  \includegraphics{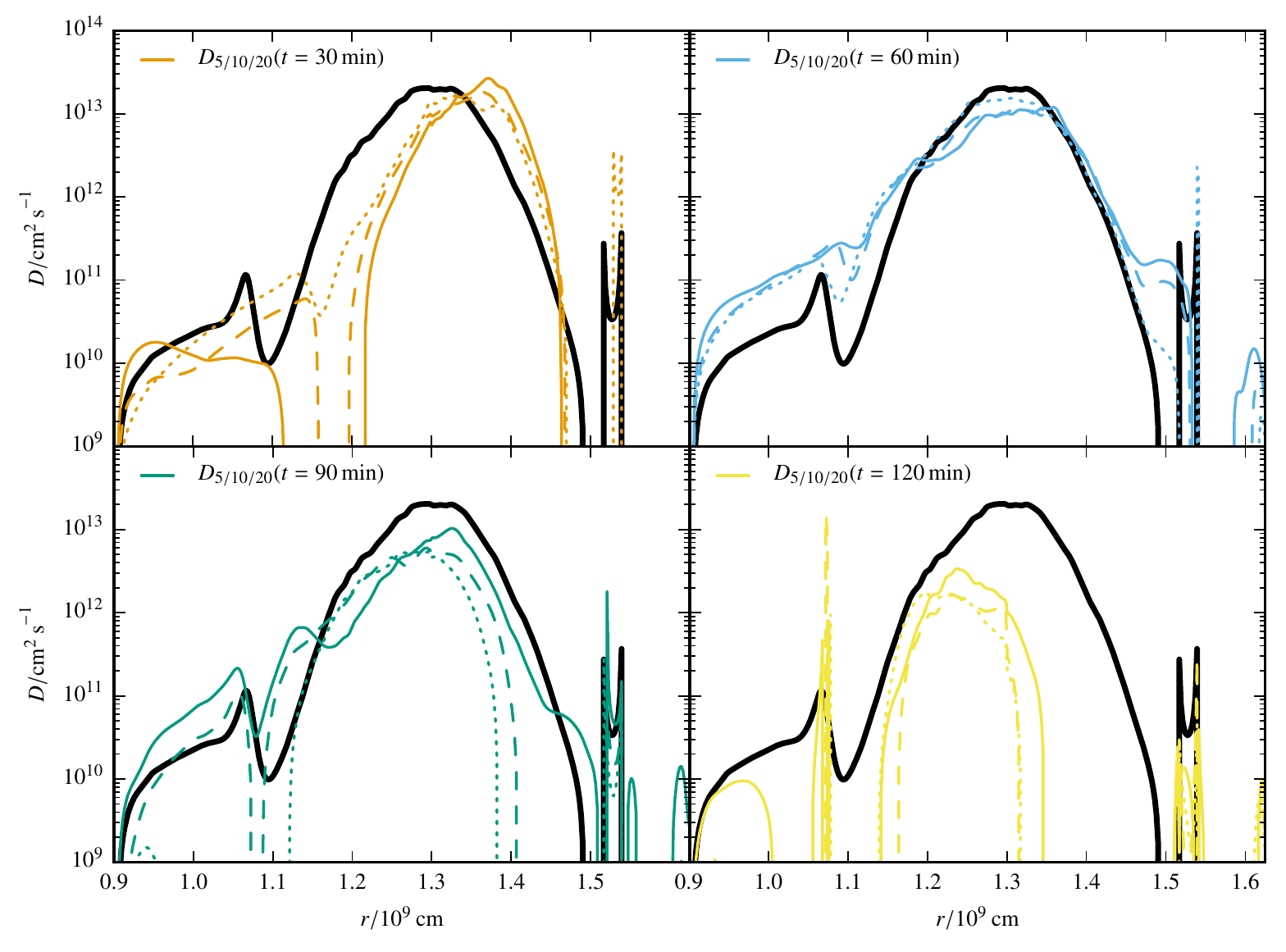}
  \caption{\label{fig:D-steps}Reconstructed diffusion coefficients calculated at
  different times during the simulation. The coefficients $D_5$ (solid), $D_{10}$
  (dashed) and $D_{20}$ (dotted) use a step size of 5~min, 10~min and 20~min,
  respectively, for the calculation. $D_{120}$ (solid black) is the same as the
  diffusion coefficient in \cref{fig:Dshear-2D}, i.e.\ using a step size of
  120~min.}
\end{figure*}

Some regions can show antidiffusive behavior, that is $D<0$. The diffusion
coefficient was set to zero in these cases. To reduce artifacts introduced by
starting the computation in a region with a flat profile, we started from both
sides and matched the results at the peak of $D$ ($r=1.3\times 10^9\,\text{cm}$).
The result is shown in \cref{fig:Dshear-2D}. As a benchmark of the quality of
the diffusion approximation in this situation we plot $\bar{A}^\text{rec}$, the
mean atomic mass profile recovered by applying $D$ to the initial profile, as
well. It is virtually identical to the final profile $\bar{A}^\text{fin}$,
which suggests that diffusion is able to reproduce the horizontally
averaged profile adequately. To check how strongly $D$ is influenced by the
choice of step size we compute diffusion coefficients for a range of smaller
step sizes (5~min, 10~min, 20~min). \Cref{fig:D-steps} shows these at different
times during the simulation. We notice that the step size does not have
a strong impact on the diffusion coefficient, at most a factor of about 2.

It is not reasonable to compare this to the shear diffusion coefficient
of GENEC at any particular instant because usually only single zones
have an extremely large ($D \approx 10^{14}\,\text{cm}^2\,\text{s}^{-1}$) value
while the coefficient in neighboring zones is zero. This behavior is an
artifact of the numerical implementation, in which only unstable zones
have a nonzero diffusion coefficient. The form of the diffusion
coefficient reconstructed from the hydro simulation underlines the fact
that a non-zero diffusion coefficient should be applied to a spatial
extent larger than the unstable zones.

\section{Conclusions}\label{sec:conclusions}
We performed a 2D~hydrodynamic simulation of a dynamical shear
instability in a massive star to study their development and the extent
of mixing they cause. To start from realistic conditions we use an
initial model from the GENEC stellar evolution code. In accordance with
linear stability analysis we observe a growing shear instability in
regions where $\Ri<0.25$. \Cref{fig:profile-lines} shows that mixing
causes the angular velocity profile to become flatter at the center and
steeper at the boundaries of the unstable layer. This shifts the
unstable region to lower radii until a stable, cylindrically symmetric
state is reached. An interesting finding of this work is that these
stable regions are in the range $0.25<\Ri<1$, despite the fact that the
flow here was initially turbulent. This suggests a critical Richardson
number of 0.25 for triggering the instability, at least in the
considered case, in which no other instabilities are present.

We compare the outcome to the treatment in the 1D~stellar
evolution code GENEC, which also provided the initial conditions. In
addition to the detailed time evolution of the profiles in
\cref{fig:ri-contour,fig:omega-contour}, we compute an effective
diffusion coefficient from the hydro simulation and compare the initial
and final profiles of mean atomic mass to the 1D~model. In
\Cref{fig:Dshear-2D} we observe that the instability occurs in 2D in a
qualitatively similar fashion to the 1D~hydrostatic prescription. This
means the position of the instability, as determined by the Richardson
criterion in 1D, is almost the same. A larger quantity of material is
mixed in the 2D case. This is not particularly surprising as the value
of the diffusion coefficient is not very well constrained by simulations
or theory so far (see \cref{rcp}).

Even very detailed 2D simulations will not give final answers
to the nature of the dynamical shear instability. This is due to
the differences between 2D and 3D~turbulence. An obvious manifestation
of this is the long term stability of rolled-up vortex sheets as seen in
the first two panels of \cref{fig:ri-panels}. A typical observation in
the 3D~case is that these large-scale structures decay much more quickly
to a turbulent state. On a more fundamental level, for 2D~turbulence
kinetic energy is transported from smaller to larger scales instead of
the opposite, which is true for the 3D~case. Future work on the
hydrodynamics of the dynamical shear instability should therefore
ideally use 3D~simulations. This is not merely a question of higher
computational cost but also raises the issue of how to map a 1D~averaged
model to an oblate 3D~spheroid. Simply adhering to the definition of shellular
rotation alone can lead to inconsistent states.

The next step will be to improve the treatment of dynamical shear in 1D codes
by considering the results of multidimensional simulations presented in this
study. The main point that any 1D implementation of dynamical shear needs to
fulfill is that redistribution of angular momentum (and the associated mixing
of composition) occurs in regions where $\Ri < 0.25$ and the instability stops
once a stable configuration is reached. Another way of looking at this is that
angular momentum will be transported (and mixing will take place) until the
angular velocity gradient is shallow enough to be stable everywhere ($\Ri
> 0.25$). Finding the necessary 1D implementation is not an easy task. The
current treatment in 1D codes already includes a very efficient transport in
unstable layers. The net effect in 1D codes is to shift the steep gradient to
the next zone rather than smooth it as in the multidimensional simulations.  Using an
even higher transport coefficient will not remedy this problem. The different
outcome in 1D models is due to the fact that the evolutionary time step is much
longer than the timescale of the dynamical shear instability. The higher
coefficient found in multidimensional simulations only applies for a small fraction of
the evolutionary time step, after which the zones are stable and the coefficient
is zero. The improved prescription needs to reproduce the long-term behavior of
the instability. Potential avenues to explore are, for example, using an artificially
smaller diffusion coefficient in the unstable layer or applying the large
diffusion coefficient over a broader region by developing a prescription
combining dynamical shear with secular shear \citep[see, e.g.,][]{maeder1997a}.
These are only suggestions. The details of new prescriptions are beyond the
scope of this paper and will be the topic of further studies.

\begin{acknowledgements}
PVFE, FKR, and SJ gratefully acknowledge support from the Klaus Tschira
Foundation.  The research leading to these results has received funding
from the European Research Council under the European Union's Seventh
Framework Programme (FP/2007-2013) / ERC Grant Agreement n. 306901.  RH
acknowledges support from the World Premier International Research
Center Initiative (WPI Initiative), MEXT, Japan. The authors thank
F.~X.\ Timmes for making his code for plasma neutrino loss rates
and his equation of state publicly available. PVFE thanks R\"udiger
Pakmor for his persistent encouragement. SJ is a fellow of the Alexander
von Humboldt Foundation.
\end{acknowledgements}

\bibliographystyle{aa}
\bibliography{dynshear}

\end{document}